\documentclass[conference, twocolumn]{IEEEtran}

\usepackage[utf8]{inputenc}
\usepackage{microtype}

\usepackage{amssymb}
\usepackage{graphicx}
\usepackage[hyphens]{url}
\usepackage[inline]{enumitem}
\usepackage{balance}
\usepackage{cite}

\usepackage{eurosym}
\usepackage{enumitem}
\usepackage [autostyle, english = american]{csquotes}

\usepackage{booktabs}
\usepackage[table,xcdraw]{xcolor}

\usepackage[flushleft]{threeparttable}

\usepackage{balance}
\usepackage{dcolumn}
\usepackage{longtable,tabu}
\usepackage{threeparttablex}
\usepackage{adjustbox}
\usepackage{tabularx}

\usepackage[underline=true]{pgf-umlsd}
\usetikzlibrary{calc}
\usetikzlibrary{arrows.meta}

\usepackage{arydshln}
\usepackage{multirow}

\usepackage{pifont}
\usepackage[caption=false]{subfig}

\usepackage[inline]{trackchanges}
\addeditor{Raja}
\addeditor{Kostas}
\addeditor{James}
\usepackage[mathlines,switch]{lineno}

\begin{document}
\title{Consumer Centric Data Control, Tracking and Transparency -- A Position Paper}
\author{}
\author{\IEEEauthorblockN{James Tapsell, Raja Naeem Akram, and Konstantinos Markantonakis}
\IEEEauthorblockA{ISG-SCC, Royal Holloway, University of London, Egham, United Kingdom\\
Email: \{James.Tapsell.2015\}@live.rhul.ac.uk, \{r.n.akram, k.markantonakis\}@rhul.ac.uk}}

\maketitle

\thispagestyle{plain}
\pagestyle{plain}

\begin{abstract}
Personal data related to a user's activities, preferences and services, is considered to be a valuable commodity not only for a wide range of technology-oriented companies like Google, Amazon and Apple but also for more traditional companies like travel/transport, banking, entertainment and marketing industry. 
This has resulted in more targeted and to a great extend personalised services for individuals -- in most cases at a minimal financial cost to them. 
The operational reality upon which a user authorises companies to collect his/her personal data to receive, in return, more personalised/targeted/context-aware services and hassle-free activities (for users) is widely deployed. 
It becomes evident that the security, integrity and accessibility of the collected data are of paramount importance. These characteristics are becoming more entrenched in the era of Internet-of-Things (IoT), autonomous vehicles and seamless travel. 
In this position paper, we examine the challenges faced by both users and organisations in dealing with the Personal Identifiable Information (PII). Furthermore, we expand on the implications of the General Data Protection Regulation (GDPR) specifically for the management of the PII. Subsequently, 
we extend the discussion to future technologies, especially the IoT and integrated transport systems for better customer experience -- and their ramification on the data governance and PII management. 
Finally, we propose a framework that balances user's privacy and data control with an organisation's objective of delivering quality, targeted and efficient services to their customers using the ``collected user data''. 
This framework is referred to as ``Consumer Oriented Data Control \& Auditability'' (CODCA) and defines the technologies that are adapted to privacy concerns and legal/regulation-frameworks. 
 
\end{abstract}


\section{Introduction}
Recent technological advances have revolutionised many day-to-day tasks by making them simpler and more convenient for the general public. 
For example, internet connectivity has simplified the task of obtaining maps and travel information on mobile phones, access to transport services and paying via smart ticketing applications. 
Organisations operating in a competitive market, offering efficient and cost-effective service provisioning, rely on data and their inherent ability to predict the short- to long-term consumer trends. 
Of particular importance are the ones related to how customers use the organisational services so that these organisations can improve their offered services to both customers and organisations. 
Therefore, we realise that most consumer-service companies rely on the data related to customer behaviour and preferences for their business-critical activities. 
Hereafter, we refer to data related to a user/consumer as ``user-data", which potentially captures the personal traits, activities and additional information to identify a unique individual. 
User-data can contain PII and additional information about an individual. 
For example, anonymised data might not be regarded as a PII \cite{narayanan2010myths}, but it is still user-data that is stripped off with any identifiers to distinguish a particular individual. 
Nevertheless, in the context of this paper, user-data is defined as:

``\textit{A set of data that represents and is associated with the identity, activities and service-offerings associated with a unique individual. Whether in an identifiable (non-anonymised) or non-identifiable (anonymised) form - collected/process/shared by an organisation (or its partners) to either provide/tailor a service to the respective individual}.''

User-data is the backbone of many of the open internet services, including Google Search, Facebook and WhatsApp, to name a few. Also, user-data is also collected by traditional organisations like superstore chains, transport companies and local authorities\footnote{The terms authority, local authority and government is associated with the entity that is responsible for keepings individual organisations accountable about their activities regarding user data - in a given geographical, political and national boundaries. These terms are used interchangeably in this paper}. Therefore, the assertion that user-data is fundamental to the functioning of a large number of organisations and services that we rely on would be an accurate portrayal of current practices related to the personal data. Therefore, a strict notion of privacy that organisations should have nothing to do with user-data is neither valid nor practical any more \cite{netter2013privacy}.

We, the users, have become so accustomed to the modern conveniences, in most cases available at no cost to users, that it is difficult to envision life without them. Similarly, as discussed before most of these conveniences necessitates collection, storage, processing and sharing of a certain level of user-data. Conceivably, many governments have mandated minimum protection for user-data, which organisations working with user-data have to abide by. Examples of such mandates include General Data Protection Regulation (GDPR)~\cite{eu:gdpr}, Health Insurance Portability and Accountability Act (HIPAA)~\cite{hipaa}, and many other initiatives globally \cite{Greenleaf2015}.

In this present situation, there are privacy regulations that an organisation has to abide. The way these regulations are enforced is dependent on local governments and/or respective data compliance/auditing authorities, ranging from audits to confirm they follow the regulations to just taking their commitment to it \cite{shah2015law,weiss2016us}. 

From the general public's point of view, they do not have any mechanism to independently and on-demand verify whether a particular organisation is following the data privacy regulations and the end-user agreement (between the individual user and the respective organisation). The only path is for them to enforce the policy or discover how their data is being used is via their local authorities - under specific regulations. For example, GDPR allows a user to request for `right-to-know' and `right-to-forget'. The above problem of managing, tracking and enforcing data governance policies\footnote{Set of policies that an organisation sets to abide by, usually they encompass the local data/information/user privacy regulations and any additional commitments the organisations make to their consumers.} with the advent of Big Data \cite{tene2011privacy}, Internet-of-Things (IoT) \cite{dahi2015privacy} and autonomous vehicles \cite{glancy2012privacy}, has only got more complex.

On the other hand, there is a growing current in the discussion surrounding data transparency as a potential corporate competitive advantage \cite{schnackenberg2016organizational} -- especially its application in clinical trials \cite{gale2012engaging, groves2013big} and other online services \cite{awad2006personalization}. 

Therefore, a potential way forward can be to design and develop  the tools and mechanisms to \textit{give consumers the control, traceability, management and auditing of their data}. Therefore, `empowering the users' to control and observe how respective organisations use their data. This `consumer empowerment' can be feasible from technologically, commercial and regulatory aspects, as the overarching commitment from both the commercial organisations and local authorities (governments) is to facilitate/serve the general public. Therefore, based on the proposed model in this paper will enable a open, fair and transparent consumer data management practices -- building trust in the digital technologies and empowering the general public (users). 

The discussion about whether such a proposal makes commercial sense is beyond the scope of this paper and we assume for our discussions that both the organisations and authorities have a single principle: consumer-first.

The main contributions of this paper can be summed up as below:
\begin{enumerate}
\item A discussion on the challenges of data ownership and control, and how it can be transferred to individual users to own/manage their data (Section \ref{sec:DataOwnershipandControlChallegnes}).
\item A framework that brings together the three main stakeholders (users, organisations, governments) to build a Consumer Data Control and Data Auditability (CODCA) framework (Section \ref{sec:CODCA}).
\item A detail discussion of building blocks of CODCA (Sections \ref{sec:ConsumerOrientedDataControl} and \ref{sec:ConsumerCentricAuditabilityFramework}).
\end{enumerate}

\section{Data Ownership and Control Challenges}
\label{sec:DataOwnershipandControlChallegnes}
In this section, we examine the data ownership and control challenges, and what they entail by taking into consideration the organisational and individual-users perspectives. Finally, we review the relationships of data triad: user, organisation and legislative-authority (government) in formulating a fair policy with enforcement mechanisms.

\subsection{Data Ownership and Control}
\label{sec:DataOwnershipandControl}
Data in the information technology domain can be defined as \emph{``a collection of numerical values (i.e. binary values) that make a uniquely identifiable set, representing a passive entity. 
Collection of data requires an active entity (i.e. software, and hardware modules) to collect, process, and communicate it''} \cite{akram2014unified}. The concept of data ownership is one of the most contentious issues related to modern digital life. 

When a user fills an online form with her details, who owns this information? In most cases, this information is probably under the control of the organisation that is collecting the information. However, do they own the data, even when it belongs to the users? What role do ``terms and conditions'' play in such collection of data? Example of different ownership perspectives can be found from Instagram \cite{McCullagh2012,McCullagh2012a} and Google privacy statements \cite{akram2014unified} -- similar to other online service providers.

Data ownership is defined as ``the legal rights and complete control over a single piece or set of data elements\footnote{Web link: \url{http://www.techopedia.com/definition/29059/data-ownership}}''. Legal rights are difficult to manage and contest in a court of law by individual consumers. The definition is closer to the organisational ownership of consumer data, not an individual's ownership of her data, although legal rights and strong accountability are proposed as a possible solution to the lack of effectiveness of traditional data security measures \cite{Weitzner2008}. From a data security point of view, data ownership is the ability to control the access, modification, and transmission of data along with the ability to track and audit the data and associated processes.

The data owner, whether it is an individual user or an organisation, has complete control over how, where, and by whom their data can be accessed. Also, the data owner should have the ability to track the data and the processes performed on it. A data owner also can delegate the administration of its data to third parties, which act as data custodians (i.e. Google in case of Google Docs). In the current context, the data owner intrinsically transfers their ownership rights (e.g. functionalities to control the data) to the data custodians. 

The data owner has to trust the data custodian and has no `technical' means of controlling any aspect of their data beyond what is sanctioned by the data custodian. The level of accountability and enforcement in the relationship between the data owner and data custodian is dependent upon the mutual agreements and the regulatory authority's related regulations. 

\begin{figure*}[ht]
 \centering
  \includegraphics[width=0.99\textwidth]{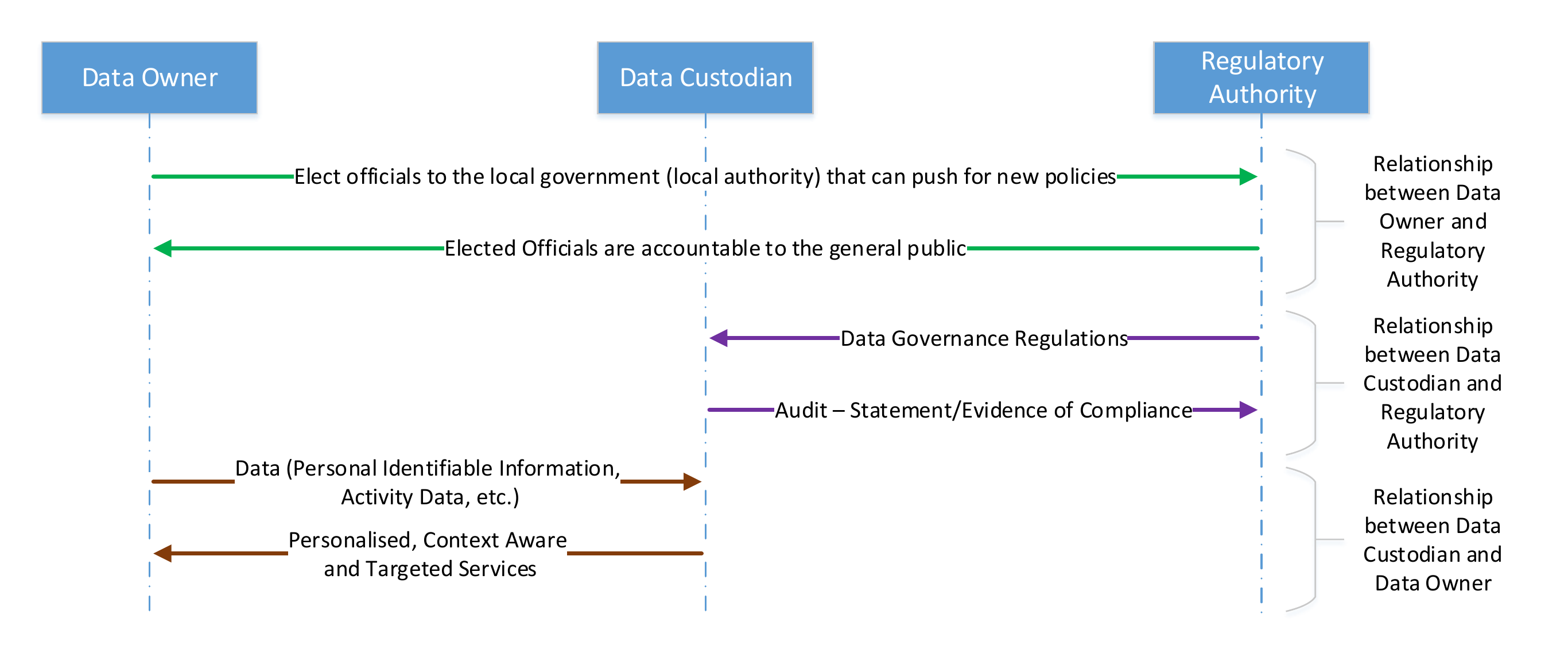}
 \caption{Relationship structure between data owner, custodian and regulatory authority.}
 \label{fig:DataRelationship}
\end{figure*}

In the context of the data users, they are individuals or organisations that utilise the data (after explicit or implicit permission from either the data owner or custodian). 
In most cases, the data owner and user are the same entity -- with few exceptions like news websites, blogs and video sharing platforms like youtube. Whereas the regulatory authority is an entity that defines and enforce data governance related policies for data custodians. 
Although it represents an over-generalisation of the different roles played by various actors in the IT infrastructure, we have restricted them to distinct categories to make it less complicated to understand the subsequent discussion.

As shown in Figure \ref{fig:DataRelationship}, data custodians get into an agreement with the data owner to access/collect and use their data to provide them better services. The data owner has to accept the agreement (also referred to as a contract) if she wants to get the services -- in some cases, she cannot gain access to services without signing the contract. One thing to note is that the contract has to be fair and can not violate any data governance regulations set by the local authorities (governments).
The regulatory authority (government) works for the betterment of the general public. Therefore, they make the data governance regulations that the data custodian has to abide by. 
The regulatory authority enforces the government regulations and if there are any violations, can even penalise the data custodians.

In this framework, the only assurance the data owner has that her data is used as per agreement and governance regulations is how effective the regulatory authority is in enforcing and auditing the data custodians. There is no independent way for the data owner to verify on-demand that their data is used as per the stated policy of the data custodian. 

To increase the user's rights regarding their data privacy and facilitate organisations to manage data more securely and transparently, the EU has pushed for the GDPR that will come in to force in May 2018. Some of the salient features of the GDPR are listed below:

\begin{itemize}
\item Privacy notices and T\&Cs must be transparent: The agreements from respective organisations that a user accepts has to be unambiguous, clear and written in plain language. Furthermore, consent to collect data has to be a `clear affirmative action' and `silence, pre-ticked boxes or inactivity' will not be considered. 
\item Consumer rights must be upheld and publicised: An organisation must take into account consumer rights when conducting `Privacy Impact Assessment' for each new process that would process the user's data. 
\item Right of subject access: Users have the right to request any organisations to provide them will all the data they have about the respective individual. This right has to be served within a reasonable time at no to minimal cost. Government organisations have the power to deny this request on national security grounds.  
\item Data portability: User data stored in an organisation should be stored in a form such that it can be moved to any other organisations if the respective users requests. 
\item Right to be forgotten: An organisation has to remove all relevant data related to a user if the respective user requests the organisation to remove the data from their storage. 
\end{itemize}

As per the GDPR, for a smaller offence, an organisation can be fined up to \euro10 million or two percent of a firm's global turnover (whichever is greater). Whereas for serious offences, an organisation can be fined up to \euro20 million or four percent of a firm's global turnover (whichever is greater) \cite{eu:gdpr}. 

In the next section, we discuss the CODCA framework to push forward the user empowerment.

\section{Consumer Oriented Data Control \& Auditability (CODCA)}
\label{sec:CODCA}
As discussed before, enabling the consumers with the ability and tools to verify that their data is being used as per the agreement with an organisation and data regulations  is the main goal of CODCA. In subsequent sections, we discuss the overall framework of the CODCA and its parts.

\subsection{Overall CODCA Framework}
\label{sec:ConsumerOrientedDataControl}

\begin{figure*}[ht]
 \centering
  \includegraphics[width=0.80\textwidth]{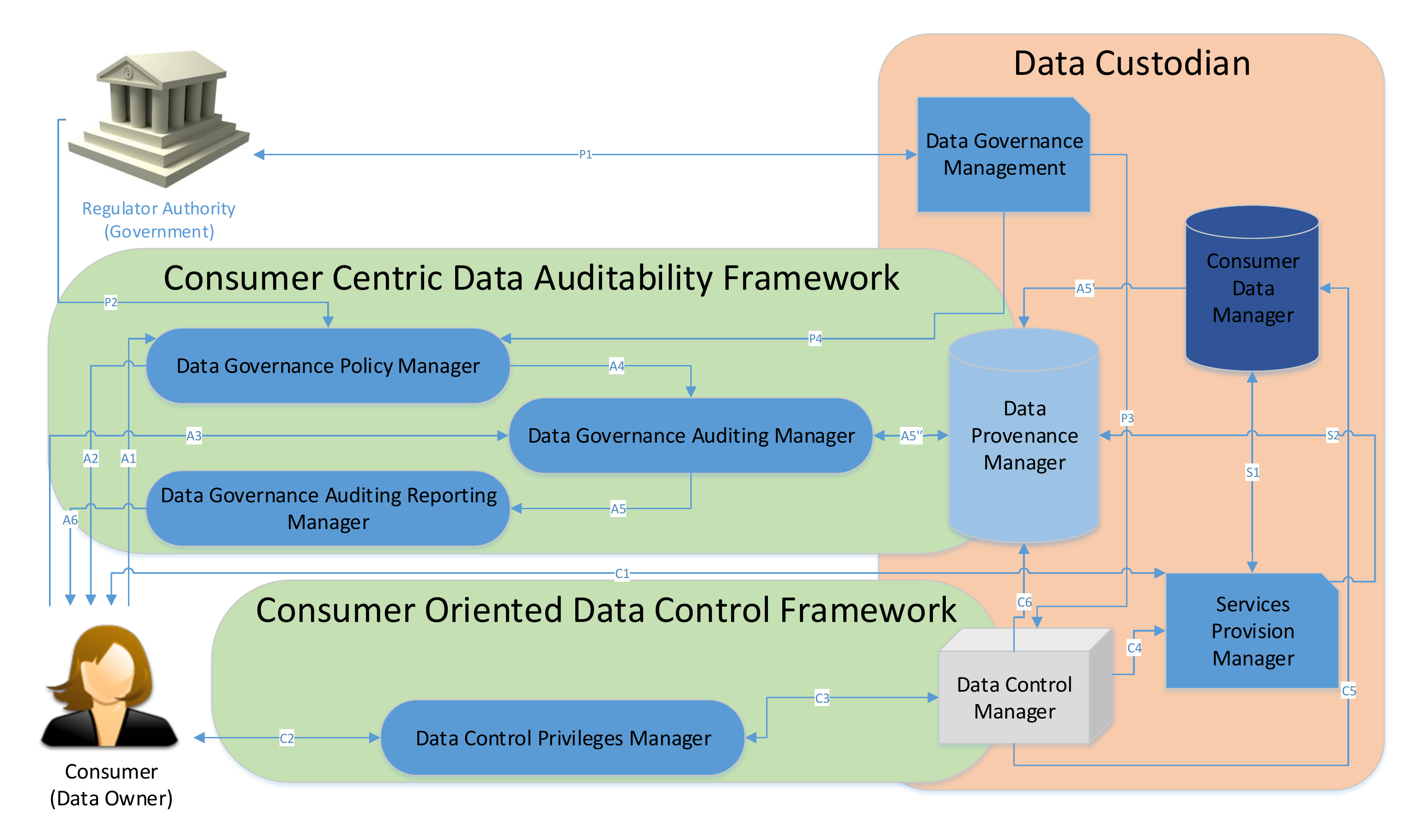}
 \caption{Overview of the CODCA Framework.}
 \label{fig:CODCA_Framework}
\end{figure*}

A degree of cooperation is required for any framework to be successful that aims to build a transparent and robust relationship between a consumer and data custodian (organisation). The same is true for the CODCA framework. Figure \ref{fig:CODCA_Framework} illustrates how different functions come together to create the CODCA framework. 

A data custodian in a generic architecture has two main components. A data store that acts as a repository for the data collected from their respective consumers and service delivery records referred to Consumer Data Manager. The second component is the actual service that is provisioned to the consumers - referred to as Service Provision Manager. 

Besides the data and provision managers, two more elements are supported by the data custodians: Data Provenance Manager and Data Control Manager. The provenance manager collects the provenance records from the entire data store of a data custodian, compiling a detailed activity record. The control manager enables a consumer to exercise their privileges as defined by the data custodian, regulatory authority and any agreements with the data  owner.

The two main components of the CODCA: Consumer-Oriented Data Control (CODC) framework and Consumer Oriented Data Auditability (CODA) framework, can either be implemented by the data custodian themselves, a third party or regulatory authority. The arrows in Figure \ref{fig:CODCA_Framework} represent communication lines and numbering on them does not describe any particular sequence of events.

The regulatory authority would set a minimum data governance requirement; most organisations might try to meet this whereas others might have some additional policies -- to show their commitment to their consumers and differentiate themselves from their competitors. The data governance requirements (P1) feed into the data governance management of organisations. This becomes an integral part of the organisation's T\&Cs and privacy policy. The T\&Cs and privacy policy are feed into the Data Governance Policy component of C2DC (P4) and Data Control Manager (P3). This defines the basic functionality and privileges a consumer would have under the agreement they sign with the organisations. Once the agreement is signed, the consumer can access the organisation's services (C1). 

The service provision manager of the organisation collects data from the user and also retrieves user-data from the internal repositories (S1) - to create a context-aware and targeted service provisioning. All activities performed by the service provision manager are recorded by the data provenance manager (S2). Data provenance manager also records events from the consumer data manager (A5'). 

On request, if a consumer wants to verify that her data is used as per the agreement between her and the organisation (A3). The Data Governance Auditing Manager would fetch the agreement (A4) and provenance records (with validity proof) from data provenance manager (A5''). The function of the auditing manager is verify whether the activities recorded by the provenance manager are in conformance with the signed agreement. The output of this analysis is fed to the reporting manager (A5), that displays the result as desired by the user -- from a simple overview to a detailed analysis.

The data control privileges manager would enable a user to request any changes to their data and exercise data control privileges\footnote{ The data governance policy stipulates the type of privileges a user can exercise, as sanctioned by the regulatory authority and the data custodian (organisation).} (C2 \& C3). 
For example, if a user requests to delete all of the data related to her. The data control privilege manager would check whether she has such a privilege or not. 
If she does, then on receiving this command the data control manager would carry out the task to remove her data and respond with an assurance that all data is deleted. 
The description in this section gives an overview of the CODCA, which can neither be implemented or successfully deployed without each entity, primarily the data custodian. 
In the next sections, we will look at how each of the entities shown in Figure \ref{fig:DataRelationship} come together collaboratively to provide CODCA.

\subsection{Consumer Oriented Data Control (CODC) Framework}
\label{sec:ConsumerOrientedDataFramework}
In this section, we discuss the role played by the data custodian, regulatory authority and consumers in the CODA framework. 

\subsubsection{Data Custodian's Role}
\label{sec:DataCollectorsRequirements}
The most prominent role in the CODCA is played by the data custodian (DC). The rationale behind the CODCA is to enable data custodians to be more transparent and provide an open platform for their consumers to validate their data commitments as stipulated by the T\&Cs and privacy policy to them. Therefore, their role is divided into two elements: responsibilities and requirements\footnote{The numbering style for the responsibility and requirements for each entity are [Entity Acroymn]-[Framework Identifier][R or Q][Num]. Responsibilities are identified as `R' and requirements as `Q'. Framework identifiers are `C' for CODC, and `A' for CODA} The set of responsibilities a data custodian has in the context of the CODC are listed as below:

\begin{enumerate}[itemindent=1.0cm,labelindent=12pt,label=DC-CR\arabic*)]
\item As part of the T\&Cs and privacy policy, define the privileges that a consumer can exercise about her data. 
\item The T\&Cs and privacy policy at a minimum should meet the local authorities stipulated data regulations. 
\item Implement the data control manager to ensure that the consumer requests are acted upon. 
\item All requests made by the consumers should be recorded in the data provenance, and any management commands like data modification and delete actions should have a proof of execution. 
\item The data control manager should provide an interface to a third party and regulatory authorities to act on behalf of the consumers. 
\item Provide an open audit to assure that all the data provenance related mechanisms are efficient and they are reporting the activities accurately. 
\item Implemented provenance records attestation/validation mechanism, to build trust in the security and integrity of these records. 
\item Optional: The data custodian might implement an interface of their own to allow consumers to exercise their privileges. 
\end{enumerate}

The set of requirements that data custodians have in the CODCA are listed below:

\begin{enumerate}[itemindent=1.0cm,labelindent=12pt,label=DC-CQ\arabic*)]
\item A data custodian's business relies on the consumer data, so they would have a sufficient privilege to process and use it to provide better services. 
\item Any third party entity interfacing with the data custodian's CODCA features would have to abide by the security and operational requirements of the data custodian.
\item Users have to be securely authenticated to exercise privileges as defined by the data custodian.
\end{enumerate}

\subsubsection{Regulatory Authority's Role}
\label{sec:RegulatoryAuthorityRequirements}
A regulatory authority (RA) in many ways has the most important role. It is the entity that can enforce and also have legal repercussions for the data custodians. The set of responsibilities they have are listed as below:

\begin{enumerate}[itemindent=1.0cm,labelindent=12pt,label=RA-CR\arabic*)]
\item Specify a balance and fair data retention and management policy that protects and ensures:
\begin{enumerate}
\item Individual's data-ownership, and 
\item The organisation's need to collect, store and process it. 
\end{enumerate}
\item Define legal repercussions for the organisations if they fail to meet the specified data management regulations. 
\item Define legal penalties if the organisations do not abide by their own declared T\&Cs and privacy policies.
\item Regularly audit the processes and practice carried out by organisations for data management. 
In the context of the CODCA, this also has to include the data provenance mechanism.
\end{enumerate}

The set of requirements a regulatory authority has about the CODCA are listed below:

\begin{enumerate}[itemindent=1.0cm,labelindent=12pt,label=RA-CQ\arabic*)]
\item They have the legal and technical ability to audit the practices of the data custodian, to verify that they are in line with the stated regulations. 
\item They have the legal ability to prosecute any data custodian if it is found to be violating the stated regulations.
\end{enumerate}

\subsubsection{Consumer's Role}
\label{sec:ConsumersRequirements}
The consumer (CO) in this ecosystem does not have any particular responsibilities. Both the data custodians and regulatory authority are the ones that are there to serve in one way or the other the consumers. However, from a requirements point of view, the list is as below:

\begin{enumerate}[itemindent=1.0cm,labelindent=12pt,label=CO-CQ\arabic*)]
\item They can control, track and verify the activities performed on their data as specified by the T\&Cs and privacy policy of the data custodian. 
\item They can either perform the privileges via data custodian's interface or delegate this to a third party that acts on behalf of the customer. 
\end{enumerate}

In the next section, we discuss the auditability requirements followed by the critical elements of the framework.

\subsection{Consumer Oriented Data Auditability (CODA) Framework}
\label{sec:ConsumerCentricAuditabilityFramework}

The process of auditability in the CODCA, similar to the CODC, relies on the collaboration of the primary stakeholders: data custodians and regulatory authorities.

\subsubsection{Data Custodian's Role}
\label{sec:DataCollectorsRequirementsR}
User data is managed by the data custodian for its entire lifetime - from collection to final removal. Therefore, the most crucial role in the auditing functionality is played by the data custodian. The set of main responsibilities on the data custodian are listed as below:

\begin{enumerate}[itemindent=1.0cm,labelindent=12pt,label=DC-AR\arabic*)]
\item They have an open and unambiguous commitment to their customers for the transparency of data management practices.
\item They implement a reliable, integrity-enabled and verifiable mechanism for collection of data provenance records across their organisation -- enabling their customer to view the operations and activities their personal data goes through at different stages. 
\item Provide access to the data provenance records related to individual customers as requests (on-demand). 
\item Provide clear and concise T\&Cs and privacy policies to their consumers against which their data management practices can be audited. 

\end{enumerate}

\subsubsection{Regulatory Authority's Role}
\label{sec:DataCollectorsRequirementsA}
As discussed in the section \ref{sec:RegulatoryAuthorityRequirements}, regulatory authorities might audit the practices of the data custodian independent of the CODCA auditability framework. In a matter of fact, the trust in the data custodian's reporting mechanism, their data provenance collection mechanism and how the data is managed internally in the organisations is based on such audits. These ideally are carried out by the regulatory authority or any of their approved auditors. From the responsibility point of view, the regulatory authority is responsible for:

\begin{enumerate}[itemindent=1.0cm,labelindent=12pt,label=RA-AR\arabic*)]
\item Provide evidence that consumers can trust which validates the data practices of the data custodian. 
\item Make the data audit report accessible to the customers who rely on the regulatory authority's evidence. 
Therefore, as part of the audit, the regulatory authority can also check the integrity and validity of data custodian's reporting standard to the customers. 
\end{enumerate}

\subsubsection{Consumer's Role}
\label{sec:ConsumersRequirements}
For a consumer to be informed about their data, flexible and user-oriented techniques have to be used. As most of the consumers are not experts in the field of auditing or data processing - it is paramount that they can recognise and make informed decisions about there data usage. For this, in the auditability framework, the requirements of the consumers are as below:

\begin{enumerate}[itemindent=1.0cm,labelindent=12pt,label=CO-AQ\arabic*)]
\item They have the privilege to access the audit reports from the data custodian in a manner that makes an intuitive sense. 
\item Consumers can retrieve the data provenance and data governance policies themselves and perform an audit, or  
\item They can involve a third party\footnote{Given that consumers trusts the third party and they are competent to carry out such audits.} that on their behalf retrieves the data provenance and data governance policies from the data custodian and performs the audit.
\end{enumerate}

Intuitively showing the audit report to users is the domain of human user interface, which is crucial for the user empowerment. However, the design work related to the visualisation of such reports is a specialised domain of research, and beyond the scope of the paper. 

In the next section, we discuss the critical elements of the CODCA followed by a discussion on why CODCA proposal would be acceptable to data custodians and other stakeholders.

\subsection{Critical Elements of the Framework}
\label{sec:ConsumerCentricAuditabilityFramework}
The CODCA framework relies on some technological elements, and without these, the features of the CODCA cannot fully materialise. These elements are discussed in this section and how potentially they can build.

\subsubsection{Data Governance Policy Extraction}
\label{sec:DataGovPolExt}
For a consumer to manage, track and audit her data, the crucial element is what are her privileges and rights as specified by the agreement she has accepted about the data custodian. This information can then be used for the management of the data (Section \ref{sec:ConsumerOrientedDataFramework}) and data audit reports. The data governance policy extractions take a humanly readable agreement (signed between the consumer and data custodian) and translate into concrete data management rule sets. 

The challenge in this context is not language processing, as it has been successfully carried out in the existing works \cite{collobert2011natural,manning2014stanford,jackson2007natural,Bird:2006:NNL:1225403.1225421}. It is extracting data governance rule set in a manner that can be later used to verify whether the data custodian is abiding by its stated policies. This also requires understanding ambiguous-language in the agreements to create actionable tasks. 

Besides analysing the agreement between the data custodian and consumers, the policy extraction process also has to get the rule set from the data related regulations (from the relevant regulatory authority).

\subsubsection{Data Provenance}
\label{sec:DataProvenance}
Data provenance, as defined by \cite{Buneman2010,zhang2011cloudprovenance,Muniswamy-Reddy2006} is the meta-data of the derivation history of data. In the traditional view of provenance, it was stated that any data provenance system should have the following four components taken from \cite{Muniswamy-Reddy:2010:PC:1855511.1855526}.

\begin{enumerate}
 \item Data-coupling: Data and its provenance should be closely coupled. Therefore, any process on the data should match with a provenance record that describes the actions accurately.
 \item Multi-object: Provenance of a data item also stores the provenance of its (data) ancestors: this is necessary to track how different data items enforcement. 
 \item Data-independent: Provenance records should not be deleted with the deletion of the associated object. Therefore, the lifetime of data provenance stretches beyond the lifetime of the associated object. 
 \item Efficient: The collection and query mechanism for the provenance should be efficient enough that it does not have a severe performance effect on the host system. 
\end{enumerate}

Early provenance systems were mainly concerned with the data quality of a database, to ensure that no error crept into large and lengthy calculations. Even if such errors appeared, a provenance system allowed a query mechanism to search the source of such an error, so other possible data items that in their provenance record included the particular node which was found to be the culprit could also be adjusted to avoid the proliferation of errors produced. A substantial body of work has been conducted in the domain of data provenance related to databases \cite{Cadenhead2011}.

In recent years, the advent of cloud computing and document security across distributed systems have given a new dimension to data provenance design and requirements \cite{zhang2011cloudprovenance,trustcloud-framework,flogger,Lu:2010:SPE:1755688.1755723}. Furthermore, provenance is collected at different levels in a system and between systems \cite{zhang2011cloudprovenance}: some mechanism collects the provenance of the application layer\footnote{most if not all of the database provenance mechanisms collect provenance in this layer}. Examples of provenance mechanism collection at the system layer are HP's TrustCloud (Flogger) \cite{trustcloud-framework,flogger}, S2Logger \cite{suen2013s2logger}, DataPROVE \cite{zhang2011cloudprovenance} and PASS \cite{Muniswamy-Reddy2006}. 

There is also an increasing call for a data-centric view over the traditional system-centric view for cloud computing \cite{ko2011}. The core requirement of data provenance in the context of the CODCA is as below: 

\begin{enumerate}
  \item Interface/API that allows a consumer to query and collect the provenance records about their respective data.
 \item Security elements that need to be provided for reliable provenance:
 \item Integrity: The assurance that provenance is not forged or tampered. An extensive range of papers have been presented on how to secure the provenance record \cite{Zhang2011,DBLP:conf/sp/SwamyCH08}; however, it was difficult to find any provenance-related work that presents a mechanism to provide complete data security. 
 \item Availability and Auditability: An auditor can check the integrity and the correctness of provenance information, though how to prohibit or detect suspicious user annotation and false provenance fabricated by malware is still an open question.
 \item Confidentially: Provenance may contain sensitive information about the data it describes, or it may be sensitive information by itself. Encryption methods and access control policies for provenance are a necessity to prevent information leakage from provenance. It is challenging to ensure confidentially when inside intruders such as privileged administrators and cloud service providers are involved.
 \item Provenance data consistency: Provenance information must be consistent with the data it describes. Inconsistency in provenance and its data can mislead both customers and service providers.
 \item Data independent persistence, also referred to as long-term persistence: A provenance system retains an object's provenance even after the object is removed. Although an object is removed, its provenance must still be present in the provenance DAG as some other objects' ancestor; deleting the object's provenance will make the DAG disconnected. An object's provenance can be removed if it has no descendants.
 \item Efficient query: The primary use of provenance data is for users to check the lineage properties of a corresponding object of interest, through external queries.  Considering the graph structure of provenance and the large size of the cloud and the objects stored in it, the efficiency of querying affects the value of provenance directly.
\end{enumerate}

Data provenance, if implemented as a light-weight mechanism at a system level, can provide an excellent auditing tool, which in our opinion is an essential component of CODCA. Another challenge the data provenance has to overcome is how it provides privacy protection while keeping the association with the data after it is being anonymised.

\subsubsection{Data Governance Audit}
\label{sec:DataGovAudit}

This process is dependent upon the previous two processes discussed in this section.  The primary challenge is to able to take the data policy rule set and data provenance and detect whether the data custodian has violated the stated policy and regulations.  On the surface, this seems like a straightforward task; query the data provenance for violations and based on the response develop the report. However, the challenge is how to manage different levels of data provenance details - as different data custodians might collect and maintain data provenance differently. 

\section{Why Consumer Oriented Data Control and Auditability?}
\label{sec:WhyConsumerOrientedDataControl}
In this section, we discuss the rationale behind the CODCA and how it can help data custodians to be more transparent and gain the trust of their customer. The three main improvement points include trustworthiness, openness and transparency, and consumer empowerment; discussed in subsequent sections.

\subsection{Integrated Services - Trustworthiness}
\label{sec:IntegratedServices-Trustwrothiness}
There is a growing trend of integrating services among both homogeneous and heterogeneous organisations. The win-win strategy is to collaborate, even with the potential competitor in the industry. This has enabled many successful services in different industries, for example banking industry in building a card-based transaction architecture - whether via the Automated Teller Machine (ATM) or POS (Point of Sales). 

For the success of such an integrated service, it is necessary to establish/build trust in individual entities. An integration that is based on data, where data collected by one entity (i.e. data custodian) is crucial to the functions of another entity, trustworthiness of not only the entity but also the data is fundamental. For example, in a multi-party train infrastructure where platforms are management by one company and trains by another. Both of these companies collected data about users and shared this with each other for their efficient operations. 

To build a framework in which a user can track and manage their data collection/usage in a system, both non-technology and technology-based trust will be needed, as both of these are complementary to each other. 

Therefore, CODCA enables the technological base of trust that a user can gain that an organisation is managing their data and organisations can trust in the validity of their data too. 

\subsection{Corporate Reputation - Openness and transparency}
\label{sec:CorporatRepuration}

Data is the foundation on which large businesses are built, especially in the context of the Internet services. Examples of such businesses include Google, Facebook and Amazon. 

Taking this with the growing number of business heavily collecting user data and who rely on this data for their day to day business along with the user's wariness of trusting organisations - due to bad news stories of data leakage, which create a desire/need for more openness and transparency. 

The CODCA is a proposal to work towards providing the most robust possible openness and transparency to the individual users, how their data is collected and used by an organisation. 

Regardless of whether the organisation is still abiding the agreement under which they collected the user data in the first place, this requirement places a level of checks \& balances in the environment which is heavily tilted towards the large organisations, and where individual users feel powerless.

\subsection{Empowerment - Consumers, the Driving Force}
\label{sec:Empowerment}
A major component of any successful business is consumer loyalty. Without consumers, in an open market, the success of a business is near to impossible.  
Therefore, in an open business environment, empowering users always generates increased consumer loyalty and trust - a cornerstone of building an enduring organisation. 
In the same spirit, CODCA empowers the consumers of organisations, the potential features which are more common and attractive to the millennial and digital generations. 

Unfortunately, there are no empirical studies related to the financial benefits of increased transparency and consumer empowerment. However, this is a growing trend in open banking like Monzo and Sterling in the UK and algorithm-less social media Vero. Therefore, from a technological point of view, we can make the recommendations that for user empowerment the CODCA framework can facilitate - however, whether it is financially beneficial or not is beyond the scope of this paper.

\section{Fundamental Rethinking Data Management}
\label{sec:FundamentRethinkingDataManagement}

In the recent light of Cambridge Analytica\footnote{URL: \url{https://www.theguardian.com/uk-news/2018/mar/22/cambridge-analytica-scandal-the-biggest-revelations-so-far}} revelations are not surprising in many respects. The scope of data collection in the Snowden revelations was substantially at larger scale then any before and since. However, the difference is data being collected and potentially manipulated by a private owned organisation that is raising some trust issues. From the Facebook's point of view the news is damaging, while it is struggling to keep its top social media platform status when faced with major challenges from mobile handset based platforms like Instagram, SnapChat and WeChat etc. 

With the GDPR regulation coming into enforcement in May 2018, organisation has to be more responsible for the data they handle. In such a situation, a potential option to gain competitive advantage is to provide data transparency as a service to their respective users. This can bridge the trust deficit currently being increasing in the general public. We do understand that at this current stage, public is not at the tipping point. However, similar news and revelations about tech giants will one day eventual push the public from where they have to actively push back the privacy advances. Before reaching such situation, best option is to go full data transparent -- enabling users the fully view of what data an organisation retains and how it is being used. In this paper, we have provide a conceptual foundation for a such a framework that with current technological advancement can be easily designed and deployed. 

Furthermore, data transparency management framework can be an enabling factor towards data monetisation. Opening up the options to individuals to retain control of their personal data and also if they do share the data, they benefit from it both from the service point of view and financially. Data transparency will also combat the fear of large organisations like Google, Facebook and Amazon collection of personal data and processing of it. 

\section{Conclusion and Future Research Directions}
\label{sec:ConclusionandFutureResearchDirections}

Consumer data is a valuable asset that is now being recognised by traditional companies like travel/transport, banking, entertainment and marketing industry. 
This trend will lead to better, targeted and personalised services for individuals -- in most cases at no or minimal financial cost to them. 
Such a  relationship is going to become more entrenched in the era of Internet-of-Things (IoT), autonomous vehicles and seamless travel. 
In this position paper, we examined the challenges faced by both the users and organisation in dealing with the Personal Identifiable Information (PII). 
We extend the discussion to the implication of the GDPR like regulations on the data management practices and the substantially large sums of fines associated with the failure to comply by the GDPR. 

The proposed framework in the paper, CODCA, provides a balance between the user's privacy (and potential desire of control) and organisations objective of delivering quality, targeted and efficient services to their customers using the user data). 
The paper advocates that empowering users can not only help organisations in making sure that they comply with the GDPR but also in building a brand image of transparency and openness. 
As the general public realises the consequences of unchecked/uncontrolled data collection and then data breaches that cause harm or inconvenience to users. 
The millennial and digital generation now look for different services models that provide more openness and consumer-participation. 
The proposed framework in this paper provides a blueprint for building services that would give control of user-data to users -- with associated technologies so they can audit and control their data usage easily. Such a service, at least from transparency and openness point of view can have a positive impact on both consumers and organisations. 
Subsequently, the paper discussed the rationality of this proposal and why commercial organisations should adopt it.

As a future research direction, many open questions 
 require resolution before CODCA framework can be realised:

\begin{itemize}
\item Data Governance Policy Extraction: Translating a text-based data collection, retention and management policy into simple statements that a user can comprehend is a challenge. Furthermore, to devise data management policies out of the data regulations (like GDPR), organisation data management policies and end-user agreements would require translation from text to actionable rule-sets. This is useful not only in apply policies enterprise-wide but also for compliance auditing. 
\item Data Provenance/Lineage Collection and Storage: Enterprise-wide collection of events in a search format will help greatly in the collection-time analysis of the events. 
\item Data Auditing Manager: Taking into account the rule sets based on relevant regulations and agreements, the auditing manager analyse the events to check if they are in compliance or not. The data auditing mechanism has to be efficient enough to provide results to all stakeholders. 
\item Data Auditing Reporting to User: Visualisation of compliance report and associated detailed activities carried on by a user is of paramount importance and a major challenge. The investigation that involves computer visualisations with human behaviour psychology and social norms that define how individual consumer signals would be essentials for the success of such a work. 
\end{itemize}




\bibliographystyle{IEEEtran}
\bibliography{references}

\begin{thebibliography}{10}
\providecommand{\url}[1]{#1}
\csname url@samestyle\endcsname
\providecommand{\newblock}{\relax}
\providecommand{\bibinfo}[2]{#2}
\providecommand{\BIBentrySTDinterwordspacing}{\spaceskip=0pt\relax}
\providecommand{\BIBentryALTinterwordstretchfactor}{4}
\providecommand{\BIBentryALTinterwordspacing}{\spaceskip=\fontdimen2\font plus
\BIBentryALTinterwordstretchfactor\fontdimen3\font minus
  \fontdimen4\font\relax}
\providecommand{\BIBforeignlanguage}[2]{{%
\expandafter\ifx\csname l@#1\endcsname\relax
\typeout{** WARNING: IEEEtran.bst: No hyphenation pattern has been}%
\typeout{** loaded for the language `#1'. Using the pattern for}%
\typeout{** the default language instead.}%
\else
\language=\csname l@#1\endcsname
\fi
#2}}
\providecommand{\BIBdecl}{\relax}
\BIBdecl

\bibitem{narayanan2010myths}
A.~Narayanan and V.~Shmatikov, ``Myths and fallacies of personally identifiable
  information,'' \emph{Communications of the ACM}, vol.~53, no.~6, pp. 24--26,
  2010.

\bibitem{netter2013privacy}
M.~Netter, M.~Riesner, M.~Weber, and G.~Pernul, ``Privacy settings in online
  social networks--preferences, perception, and reality,'' in \emph{System
  Sciences (HICSS), 2013 46th Hawaii International Conference on}.\hskip 1em
  plus 0.5em minus 0.4em\relax IEEE, 2013, pp. 3219--3228.

\bibitem{eu:gdpr}
\BIBentryALTinterwordspacing
``{Regulation (EU) 2016/679 of the European Parliament and of the Council of 27
  April 2016 on the protection of natural persons with regard to the processing
  of personal data and on the free movement of such data, and repealing
  Directive 95/46/EC (General Data Protection Regulation)},'' \emph{Official
  Journal of the European Union}, vol. L119/59, May 2016. [Online]. Available:
  \url{http://eur-lex.europa.eu/legal-content/EN/TXT/?uri=OJ:L:2016:119:TOC}
\BIBentrySTDinterwordspacing

\bibitem{hipaa}
{Centers for Medicare \& Medicaid Services}, ``{The Health Insurance
  Portability and Accountability Act of 1996 (HIPAA)},'' Online at
  http://www.cms.hhs.gov/hipaa/, 1996.

\bibitem{Greenleaf2015}
G.~Greenleaf, ``Global data privacy laws 2015: 109 countries, with european
  laws now a minority,'' \emph{Privacy Laws \& Business International Report,},
  vol. 133, no. UNSW Law Research Paper No. 2015-21, p.~7, February 2015.

\bibitem{shah2015law}
R.~Shah, ``Law enforcement and data privacy-a forward-looking approach,''
  \emph{Yale LJ}, vol. 125, p. 543, 2015.

\bibitem{weiss2016us}
M.~A. Weiss and K.~Archick, ``Us-eu data privacy: from safe harbor to privacy
  shield,'' 2016.

\bibitem{tene2011privacy}
O.~Tene and J.~Polonetsky, ``Privacy in the age of big data: a time for big
  decisions,'' \emph{Stan. L. Rev. Online}, vol.~64, p.~63, 2011.

\bibitem{dahi2015privacy}
A.~Dahi, ``Privacy and security in a connected world – the {US-FTC} on the
  internet of things,'' \emph{ZD-aktuell 2015}, p. 04569, 2015.

\bibitem{glancy2012privacy}
D.~J. Glancy, ``Privacy in autonomous vehicles,'' \emph{Santa Clara L. Rev.},
  vol.~52, p. 1171, 2012.

\bibitem{schnackenberg2016organizational}
A.~K. Schnackenberg and E.~C. Tomlinson, ``Organizational transparency: A new
  perspective on managing trust in organization-stakeholder relationships,''
  \emph{Journal of Management}, vol.~42, no.~7, pp. 1784--1810, 2016.

\bibitem{gale2012engaging}
C.~P. Gale, C.~Weston, S.~Denaxas, D.~Cunningham, M.~A. de~Belder, H.~H. Gray,
  R.~Boyle, J.~E. Deanfield \emph{et~al.}, ``Engaging with the clinical data
  transparency initiative: a view from the national institute for
  cardiovascular outcomes research (nicor),'' 2012.

\bibitem{groves2013big}
P.~Groves, B.~Kayyali, D.~Knott, and S.~Van~Kuiken, ``The ‘big
  data’revolution in healthcare,'' \emph{McKinsey Quarterly}, vol.~2, p.~3,
  2013.

\bibitem{awad2006personalization}
N.~F. Awad and M.~S. Krishnan, ``The personalization privacy paradox: an
  empirical evaluation of information transparency and the willingness to be
  profiled online for personalization,'' \emph{MIS quarterly}, pp. 13--28,
  2006.

\bibitem{akram2014unified}
R.~N. Akram and R.~K. Ko, ``Unified model for data security-a position paper,''
  in \emph{Trust, Security and Privacy in Computing and Communications
  (TrustCom), 2014 IEEE 13th International Conference on}.\hskip 1em plus 0.5em
  minus 0.4em\relax IEEE, 2014, pp. 831--839.

\bibitem{McCullagh2012}
\BIBentryALTinterwordspacing
D.~McCullagh. (2012, December) \BIBforeignlanguage{English}{{Instagram says it
  now has the right to sell your photos}}. Online. CNet. [Online]. Available:
  \url{http://news.cnet.com/8301-13578_3-57559710-38/instagram-says-it-now-has-the-right-to-sell-your-photos/}
\BIBentrySTDinterwordspacing

\bibitem{McCullagh2012a}
\BIBentryALTinterwordspacing
D.~McCullagh and D.~Tam. (2012, December)
  \BIBforeignlanguage{English}{{Instagram apologizes to users: We won't sell
  you photos}}. Online. CNet. [Online]. Available:
  \url{http://news.cnet.com/8301-1023_3-57559890-93/instagram-apologizes-to-users-we-wont-sell-your-photos/}
\BIBentrySTDinterwordspacing

\bibitem{Weitzner2008}
D.~J. Weitzner, H.~Abelson, T.~Berners-Lee, J.~Feigenbaum, J.~A. Hendler, and
  G.~J. Sussman, ``{Information Accountability},'' \emph{Commun. ACM}, vol.~51,
  no.~6, pp. 82--87, 2008.

\bibitem{collobert2011natural}
R.~Collobert, J.~Weston, L.~Bottou, M.~Karlen, K.~Kavukcuoglu, and P.~Kuksa,
  ``Natural language processing (almost) from scratch,'' \emph{Journal of
  Machine Learning Research}, vol.~12, no. Aug, pp. 2493--2537, 2011.

\bibitem{manning2014stanford}
C.~D. Manning, M.~Surdeanu, J.~Bauer, J.~R. Finkel, S.~Bethard, and
  D.~McClosky, ``The stanford corenlp natural language processing toolkit.'' in
  \emph{ACL (System Demonstrations)}, 2014, pp. 55--60.

\bibitem{jackson2007natural}
P.~Jackson and I.~Moulinier, \emph{Natural language processing for online
  applications: Text retrieval, extraction and categorization}.\hskip 1em plus
  0.5em minus 0.4em\relax John Benjamins Publishing, 2007, vol.~5.

\bibitem{Bird:2006:NNL:1225403.1225421}
\BIBentryALTinterwordspacing
S.~Bird, ``Nltk: The natural language toolkit,'' in \emph{Proceedings of the
  COLING/ACL on Interactive Presentation Sessions}, ser. COLING-ACL '06.\hskip
  1em plus 0.5em minus 0.4em\relax Stroudsburg, PA, USA: Association for
  Computational Linguistics, 2006, pp. 69--72. [Online]. Available:
  \url{http://dx.doi.org/10.3115/1225403.1225421}
\BIBentrySTDinterwordspacing

\bibitem{Buneman2010}
\BIBentryALTinterwordspacing
P.~Buneman and Susan, ``{Data Provenance - the foundation of data quality},''
  September 2010. [Online]. Available:
  \url{http://www.sei.cmu.edu/measurement/research/upload/Davidson.pdf}
\BIBentrySTDinterwordspacing

\bibitem{zhang2011cloudprovenance}
O.~Q. Zhang, M.~Kirchberg, R.~K.~L. Ko, and B.~S. Lee, ``How to track your
  data: The case for cloud computing provenance,'' in \emph{Cloud Computing
  Technology and Science (CloudCom), 2011 IEEE Third International Conference
  on}.\hskip 1em plus 0.5em minus 0.4em\relax IEEE, 2011, pp. 446--453.

\bibitem{Muniswamy-Reddy2006}
\BIBentryALTinterwordspacing
K.-K. Muniswamy-Reddy, D.~A. Holland, U.~Braun, and M.~Seltzer,
  ``{Provenance-aware Storage Systems},'' in \emph{Proceedings of the Annual
  Conference on USENIX '06 Annual Technical Conference}, ser. ATEC '06.\hskip
  1em plus 0.5em minus 0.4em\relax Berkeley, CA, USA: USENIX Association, 2006,
  pp. 4--4. [Online]. Available:
  \url{http://dl.acm.org/citation.cfm?id=1267359.1267363}
\BIBentrySTDinterwordspacing

\bibitem{Muniswamy-Reddy:2010:PC:1855511.1855526}
\BIBentryALTinterwordspacing
K.-K. Muniswamy-Reddy, P.~Macko, and M.~Seltzer, ``{Provenance for the
  Cloud},'' in \emph{Proceedings of the 8th USENIX Conference on File and
  Storage Technologies}, ser. FAST'10.\hskip 1em plus 0.5em minus 0.4em\relax
  Berkeley, CA, USA: USENIX Association, 2010, pp. 15--14. [Online]. Available:
  \url{http://dl.acm.org/citation.cfm?id=1855511.1855526}
\BIBentrySTDinterwordspacing

\bibitem{Cadenhead2011}
\BIBentryALTinterwordspacing
T.~Cadenhead, M.~Kantarcioglu, and B.~Thuraisingham, ``A framework for policies
  over provenance,'' in \emph{3rd USENIX Workshop on the Theory and Practice of
  Provenance}, P.~Buneman and J.~Freire, Eds.\hskip 1em plus 0.5em minus
  0.4em\relax Crete, Greece: USENIX, June 2011. [Online]. Available:
  \url{https://www.usenix.org/legacy/event/tapp11/tech/final_files/Cadenhead.pdf}
\BIBentrySTDinterwordspacing

\bibitem{trustcloud-framework}
R.~K.~L. Ko, P.~Jagadpramana, M.~Kirchberg, Q.~Liang, M.~Mowbray, S.~Pearson,
  and B.~S. Lee, ``Trustcloud - a framework for accountability and trust in
  cloud computing,'' in \emph{IEEE 2nd Cloud Forum for Practitioners (IEEE ICFP
  2011)}.\hskip 1em plus 0.5em minus 0.4em\relax Washington DC, USA: IEEE
  Computer Society, July 2011.

\bibitem{flogger}
R.~K.~L. Ko, P.~Jagadpramana, and B.~S. Lee, ``Flogger: A file-centric logger
  for monitoring file access and transfers within cloud computing
  environments,'' in \emph{3rd IEEE International Workshop on Security in
  e-Science and e-Research (IEEE ISSR 2011), in conjunction with IEEE TrustCom
  2011}.\hskip 1em plus 0.5em minus 0.4em\relax Changsha, China: IEEE, 2011.

\bibitem{Lu:2010:SPE:1755688.1755723}
\BIBentryALTinterwordspacing
R.~Lu, X.~Lin, X.~Liang, and X.~S. Shen, ``{Secure Provenance: The Essential of
  Bread and Butter of Data Forensics in Cloud Computing},'' in
  \emph{Proceedings of the 5th ACM Symposium on Information, Computer and
  Communications Security}, ser. ASIACCS '10.\hskip 1em plus 0.5em minus
  0.4em\relax New York, NY, USA: ACM, 2010, pp. 282--292. [Online]. Available:
  \url{http://doi.acm.org/10.1145/1755688.1755723}
\BIBentrySTDinterwordspacing

\bibitem{suen2013s2logger}
C.~H. Suen, R.~K.~L. Ko, Y.~S. Tan, P.~Jagadpramana, and B.~S. Lee, ``S2logger:
  End-to-end data tracking mechanism for cloud data provenance,'' in
  \emph{Trust, Security and Privacy in Computing and Communications (TrustCom),
  2013 12th IEEE International Conference on}, 2013, pp. 594--602.

\bibitem{ko2011}
R.~K.~L. Ko, M.~Kirchberg, and B.~S. Lee, ``From system-centric to data-centric
  logging-accountability, trust \& security in cloud computing,'' in
  \emph{Defense Science Research Conference and Expo (DSR), 2011}.\hskip 1em
  plus 0.5em minus 0.4em\relax IEEE, 2011, pp. 1--4.

\bibitem{Zhang2011}
\BIBentryALTinterwordspacing
Z.~Bao and S.~B. Davidson, ``A fine-grained workflow model with
  provenance-aware security views,'' in \emph{3rd USENIX Workshop on the Theory
  and Practice of Provenance}, P.~Buneman and J.~Freire, Eds.\hskip 1em plus
  0.5em minus 0.4em\relax Crete, Greece: USENIX, June 2011. [Online].
  Available:
  \url{https://www.usenix.org/legacy/event/tapp11/tech/final_files/Bao.pdf}
\BIBentrySTDinterwordspacing

\bibitem{DBLP:conf/sp/SwamyCH08}
N.~Swamy, B.~J. Corcoran, and M.~Hicks, ``Fable: A language for enforcing
  user-defined security policies,'' in \emph{IEEE Symposium on Security and
  Privacy}, 2008, pp. 369--383.

\end{thebibliography}

\end{document}